\documentclass[journal=jacsat,manuscript=article]{achemso}

\usepackage{amsmath}
\usepackage{amsmath,amssymb}
\usepackage{graphicx}% Include figure files
\usepackage{caption}
\usepackage{color}% Include colors for document elements
\usepackage{dcolumn}% Align table columns on decimal point
\usepackage{bm}% bold math
\usepackage{float}
\usepackage{subfig}

\author{Debasish Koner}
\altaffiliation{These authors contributed equally.}
\affiliation{Department of Chemistry, University of Basel,
Klingelbergstrasse 80, 4056 Basel, Switzerland}

\author{Seyedeh Maryam Salehi}
\altaffiliation{These authors contributed equally.}
\affiliation{Department of Chemistry, University of Basel,
  Klingelbergstrasse 80, 4056 Basel, Switzerland}

\author{Padmabati Mondal} \altaffiliation{These authors contributed
  equally.}  \affiliation{Indian Institute of Science Education and
  Research (IISER) Tirupati, Karakambadi Road, Mangalam,
  Tirupati-517507, Andhra Pradesh, India.}

\author{Markus Meuwly} \email{m.meuwly@unibas.ch}
\affiliation[University of Basel] {Department of Chemistry, University
  of Basel,\\ Klingelbergstrasse 80, 4056 Basel,
  Switzerland\\Department of Chemistry, Brown University, Providence,
  RI, USA}

\date{\today}

\title{Perspective: Non-conventional Force Fields for Applications in
  Spectroscopy and Chemical Reaction Dynamics}

\begin{document}

\begin{abstract}
Extensions and improvements of empirical force fields are discussed in
view of applications to computational vibrational spectroscopy and
reactive molecular dynamics simulations. Particular focus is on
quantitative studies which make contact with experiments and provide
complementary information for a molecular-level understanding of
processes in the gas phase and in solution. Methods range from
including multipolar charge distributions to reproducing kernel
Hilbert space approaches and machine learned energy functions based on
neural networks.
\end{abstract}

\section{Introduction}
Atomistic Simulations have become a powerful way to provide
molecular-level insights into gas- and condensed-phase properties of
physico-chemical systems and to interpret and complement
experiments. Much of the progress is directly related to the
dramatically increased capabilities of modern computer
infrastructure. Nevertheless, the field is still far from solving
general problems by running either sufficiently long (quasi-)classical
trajectories or large-scale quantum dynamics simulations in order to
characterize a particular system. Even for an atom+diatom system the
comprehensive calculation of all state-to-state cross sections is a
formidable task and brute-force sampling is a serious computational
undertaking.\cite{MM.nnsig:2019,grover:2019}\\

\noindent
For studying the dynamics of complex systems, means to determine their
total energies and associated forces are required. Empirical force
fields were initially developed to characterize the chemical structure
and dynamics of macromolecules, including peptides and
proteins~\cite{lifson68,Levi69,Hagler942,Hagler941,charmm-md,amber,opls,gromos}.
Their primary applications concerned sampling conformations of large
molecules, such as proteins in the gas phase\cite{karplus:1977} and in
solution.\cite{brooks:1993} The typical mathematical form of an
empirical force field contains bonded and nonbonded terms. The bonded
part is
\begin{eqnarray}
  V_{\rm bond} &=& \sum K_b (r-r_e)^2 \nonumber \\
  V_{\rm angle} &=& \sum K_{\theta} (\theta - \theta_e)^2 \nonumber \\
  V_{\rm dihe} &=& \sum K_{\phi} (1+\cos(n \phi - \delta))
\label{eq:ffbond}
\end{eqnarray}
and is geared towards conformational sampling, spectroscopy or
thermodynamics and not, e.g., towards describing chemical reactions
for which bonds need to be broken and formed. In Eq. \ref{eq:ffbond}
the constants $K$ are the force constants associated with the
particular type of interaction, $r_e$ and $\theta_e$ are equilibrium
values, $n$ is the periodicity of the dihedral and $\delta$ is the
phase which determines the location of the maximum. The sums run over
all respective terms. Nonbonded interactions include electrostatic and
van der Waals terms
\begin{eqnarray}
V_{\rm elstat} &=& \frac{1}{4\pi\epsilon_0} \sum \frac{q_i
  q_j}{r_{ij}} \nonumber \\
V_{\rm vdW} &=& \sum \epsilon_{ij} \left [ \left (\frac{R_{{\rm
        min},ij}}{r_{ij}} \right )^{12} - 2\left (\frac{R_{{\rm
        min},ij}}{r_{ij}} \right)^6 \right ]
\label{eq:ffnonbond}
\end{eqnarray}
where the sums run over all nonbonded atom pairs. $q_i$ and $q_j$ are
the partial charges of the atoms $i$ and $j$, and $\epsilon_0$ is the
vacuum dielectric constant. For the van der Waals terms, the potential
energy is expressed as a Lennard-Jones potential with well depth
$\epsilon_{ij}=\sqrt {\epsilon_i \epsilon_j}$ and range $R_{{\rm
    min},ij} = (R_{{\rm min},i}+R_{{\rm min},j})/2$ at the
Lennard-Jones minimum, according to standard mixing
rules.\cite{nasrifar.mixing:2019} This interaction captures long-range
dispersion $( \propto -R^{-6})$ and exchange repulsion $(\propto
R^{-12})$ where the power of the latter is chosen for convenience. The
combination of Eqs.~\ref{eq:ffbond} and ~\ref{eq:ffnonbond}
constitutes a minimal model for an empirical force field.\\

\noindent
One of the desirable features of an empirical force field is that it
is a computationally efficient model that can be improved in various
ways based on physical principles such as to allow direct comparison
with experiment and eventually provide a basis for
interpretation. Forces can be evaluated analytically which is an
advantage for molecular dynamics (MD) simulations. For example,
harmonic bonds can be replaced by Morse oscillators which are either
useful to investigate chemical reactions or accurate vibrational
spectroscopy. These two aspects are of particular relevance to the
present work.\\

\noindent
Using point charges for the electrostatic interactions can be
considered as the zero-order expansion of a more comprehensive
multipolar expansion which considerably improves the description of
the charge distribution around molecular building
blocks.\cite{stone1996theory,kramer2012atomic,bereau2013,kramer2013deriving,MM.dcm:2014,MM.dcm:2017}
In addition, the standard form of an empirical force field can be
complemented by including further terms such as
polarization.\cite{mackerell:2015} The extensibility and specific ways
to improve the quality of an empirical force field make it an ideal
starting point for using them in increasingly realistic simulations of
molecular materials at an atomistic level.\\

\noindent
Following chemical reactions - i.e. the breaking and formation of
chemical bonds - as a function of time lies at the heart of
chemistry. However, the most obvious and direct route, evaluating the
electronic structure along a molecular dynamics (MD) trajectory or
time-dependent quantum dynamics simulation, is often impractical for
several reasons. For one, evaluating the total energy and forces by
solving the electronic Schr\"odinger equation is computationally
prohibitive. This precludes using high-level quantum chemistry methods
and/or sufficiently large basis sets for running a statistically
significant number of independent trajectories. For converged
observables, averages over initial conditions need to be
computed. Also, some electronic structure calculations can be affected
by systematic error, such as basis set superposition errors, and
convergence of the Hartree-Fock wavefunction to the desired state can
be challenging. This can occur for systems containing metal atoms,
higher excited states or systems with challenging electronic
structure.\\

\noindent
In the present perspective several improvements to empirical energy
functions and their application that have been pursued in the more
recent past will be highlighted. The applications focus on vibrational
spectroscopy, reactive dynamics and thermodynamic properties which
provide a wide range of challenging problems for understanding and
interpreting physico-chemical experiments at a molecular level.\\

\section{Methodological Background}
Both, bonded and non-bonded parts of an empirical force field can be
systematically improved. An instructive example are recent efforts to
correctly describe condensed-phase and spectroscopic properties of
liquid water. Building on the realization that multipolar interactions
are
required,\cite{nutt-biophys-03,amoeba:2003,iamoeba:2013,Stone05p1128,MM.mtp:2012,MM.mtp:2013,kramer2013deriving}
the TL4P water model was developed.\cite{tavan:2013} TL4P is a rigid
water model with the atoms fixed to the experimental liquid phase
geometry. The electrostatic interactions are described by three point
charges (one on each hydrogen atom and one on a massless point M that
is displaced from the oxygen atom) and a Gaussian inducible
dipole. The van der Waals interactions are represented as a Buckingham
potential acting between oxygen atoms. With this model, parametrized
to reference density functional theory calculations, a wide range of
bulk properties could be described reliably compared with experiment
(see Table 7 of Ref.\cite{tavan:2013}).\\

\noindent
When using TL4P for computing the low-frequency part of the infrared
and Raman spectra of liquid water it was found that the shapes of the
spectral features did not compare well with
experiment.\cite{bertie:1996} Based on earlier work it was decided to
augment the TL4P water model with some amount of charge transfer along
the O--HO intramolecular hydrogen bond and to include anisotropic
polarizability.\cite{MM.water:2018} The parameters were again
determined from reference DFT calculations and both, bulk properties
and the optical spectra were computed. With these modifications, the
bulk properties were in equally good agreement with experiment as for
the original TL4P model\cite{tavan:2013} whereas the relative
intensities and peak positions for both, infrared and Raman, spectra
compared well with experiment. In this way, a physically motivated,
and computationally efficient model for water was developed by
refinement, largely guided by comparison with experiment.\\

\noindent
Following very different strategies, other successful force fields for
water have been developed in the recent past. One of them is the
AMOEBA model which builds explicitly on atomic
multipoles.\cite{amoeba:2003,iamoeba:2013} Another strategy is
followed in the E3B model which is based on adding explicit three-body
terms, akin to a many body expansion.\cite{e3b:2008} Similarly, the
HBB (Huang, Braams, Bowman) force field also uses a many-body
expansion.\cite{hbb:2006} As a more recently developed force field,
the RexPoN force field is entirely based on reference electronic
structure calculations without empirical data.\cite{goddard:2018}
Finally, the most comprehensive water force field in various phases is
probably the MB-Pol model which also builds on multipolar interactions
and many-body polarization.\cite{mbpol:2013}\\

\subsection{Multipolar and Polarizable Force Fields}
Multipolar interactions have been considered early on in molecular
recognition.\cite{rein1973physical} Different from the spherically
symmetric field around a point charge, atomic multipoles can capture
anisotropic interactions. A typical example is carbon monoxide which
can not be realistically modelled with two atom-centered point charges
located each of the two atoms (Eq. \ref{eq:ffnonbond}) because the
total charge ($Q=0$) and the total molecular dipole $\mu = 0.048$
$ea_{0}$ lead to two partial charges close to zero but with opposite
sign. In order to describe the substantial molecular quadrupole
moment\cite{maroulis96,Roco01,Roco01_ii} ($\Theta = -1.58$
$ea_{0}^{2}$) either a third interaction site midway the two atoms can
be included\cite{Str91} or the two atoms are described by a
distributed multipole
expansion.\cite{stone1996theory,nutt-biophys-03,nutt-pnas-04,Plattner08}\\

\noindent
More generally, the electrostatic potential (ESP) around a molecule
can be represented as a superposition of point charges (PC) and higher
multipole (MTP) moments. Capturing strongly anisotropic
features---e.g., lone pairs, hydrogen bonding, $\pi$-electron density
or sigma holes
(halogens)\cite{Clark2007sigmahole,Murray2009expansion,Politzer2010halobonding}
requires schemes beyond point charge representations. The ESP
$\Phi({\bf r})$ is related to the charge distribution $\rho({\bf r})$
through
\begin{equation}
  \label{eqn:espint}
  4\pi\varepsilon_0 \Phi({\bf r}) = \int {\rm d}{\bf r}'
  \frac{\rho({\bf r}')}{\left|{\bf r} - {\bf r}'\right|}= \sum_{l=0}^\infty \sum_{m=-l}^l
  \frac{Q_{lm}}{r^{l+1}} \sqrt{\frac{4\pi}{2l+1}}Y_{lm}(\theta,\phi),
\end{equation}
where ${\bf r}$ and ${\bf r}'$ are spatial variables. For the second
equation $1/|{\bf r} - {\bf r}'|$ was expanded in powers of $r'/r<1$
\cite{electrodynamics1998jd} to represent the ESP as a sum over
spherical harmonics $Y_{lm}(\theta,\phi)$ from which the spherical MTP
moments $Q_{lm}$ are defined by
\begin{equation}
  \label{eqn:qlm}
  Q_{lm} = \int {\rm d}{\bf r}' \rho({\bf r}') (r')^l
  \sqrt{\frac{4\pi}{2l+1}}Y^*_{lm}(\theta',\phi')
\end{equation}
These can be determined from the density by integration. This provides
a compact atom-based representation of the ESP around a molecule.\\

\noindent
Alternatively, multipoles of a given order can be represented by
distributed point charges as is done in the distributed charge model
(DCM).\cite{MM.dcm:2014,MM.dcm:2017} Such a model replaces the
evaluation of multipole-multipole interactions at the expense of using
a larger number of terms. Reducing the number of interaction sites can
be accomplished by finding the smallest number of point charges to
accurately represent the ESP using differential
evolution.\cite{MM.dcm:2017}\\

\subsection{Reactive Force Fields}
The design of reactive force fields has followed different routes.
Early work by Penney\cite{penney:1937} and Pauling\cite{pauling47}
related bond order, bond length, and bond strength. The finding that
bond order and bond length follow a near linear
relation\cite{pauling47} and a log-log plot of dissociation energies
versus bond order is also almost linear was used to develop ``bond
energy bond order'' (BEBO) potentials.\cite{johnston63} One of the
essential assumptions underlying this approach is that - at least for
hydrogen-atom transfer reactions - the sum of the bond orders of the
broken bond, $n_1$, and of the newly formed bond, $n_2$, fulfills $n_1
+ n_2 =1$. In other words: "At all stages of the reaction the
formation of the second bond must be 'paying for' the breaking of the
first bond''.\cite{johnston63}\\

\noindent
An extension of the concept underlying BEBO is
ReaxFF.\cite{goddard01reaxff} Here, nonbonded terms are included from
the beginning and the dissociation and reaction curves are derived
from fitting to electronic structure
calculations.\cite{goddard01reaxff,reaxff:2008} Central to ReaxFF is
that the bond order is related to the distance between two atoms. As
an example, for two carbon atoms there can be anything between ``no
bond'' (bond order = 0) to a triple bond. In ReaxFF the total energy
consists of a sum of individual terms such as $E_{\rm bond}$, a
correction for over-coordination, a penalty term $E_{\rm over}$ for
under-coordinated atoms reflecting resonance energies between
$\pi-$electrons, a conjugation energy and others. Contrary to
empirical force fields\cite{charmm-md,amber,opls}, ReaxFF does not
build on the concept of atom types. While the functional form of
ReaxFF is universal, its application to concrete problems always
involves more or less extensive fitting to reference
calculations.\cite{reaxff:2016}\\

\noindent
Chemical reactivity can be encapsulated in a natural fashion within
the framework of valence bond theory. For ionic bond cleavage
AB$\rightarrow$A$^-$+B$^+$, three chemically relevant states are
considered within empirical valence bond (EVB)\cite{warshel80evb}
theory: $\psi_1=AB$, $\psi_2=A^-B^+$, and $\psi_3=A^+B^-$. If A is
more electronegative than B (i.e. formation B$^-$ is unlikely in the
presence of A), resonance structure $\psi_3$ is largely irrelevant and
the process can be described by $\psi_1$ and $\psi_2$ alone. For a
more general chemical reaction, one must choose, ``..on the basis of
experience and intuition, a set of bonding arrangements or ``resonance
structures''[..]'' as was noted in the original EVB
publication.\cite{warshel80evb} The choice of the states in EVB is not
always obvious {\it a priori} but may also need to be based, e.g., on
the requirement to best reproduce the reference electronic structure
calculations.\cite{glowacki2015}\\

\noindent
The total $n-$state EVB Hamiltonian $H_{nn}$ contains diagonal terms
(force fields for each of the $n$ states) and the coupling between
them ($H_{ij}$). Representation and parametrization of the
off-diagonal terms has been a source of considerable
discussion,\cite{truhlar09evb,truhlar09evberr,warshel09evb} in
particular the assumption that upon transfer of the reaction from the
gas phase to the solution phase these elements do not change
significantly.\cite{warshel06evb} Applications of EVB include
enzymatic reactions (for which it was originally
developed~\cite{warshel84pnas}), proton transfer processes, and the
autodissociation of water~\cite{warshel02}. More recently, this has
been extended to other types of reactions, including bimolecular
reactions\cite{orrewing:2011,glowacki:2011} or the association and
dissociation of CH$_3$ from diamond surfaces.\cite{glowacki:2016}
Furthermore, several extensions have been suggested to the original
EVB method allowing its application to a wider class of
problems.\cite{miller90evb,voth:1998.evb,schlegel:2006}.\\

\begin{figure}[th]
\subfloat[]{\includegraphics[width = 3.40 in]{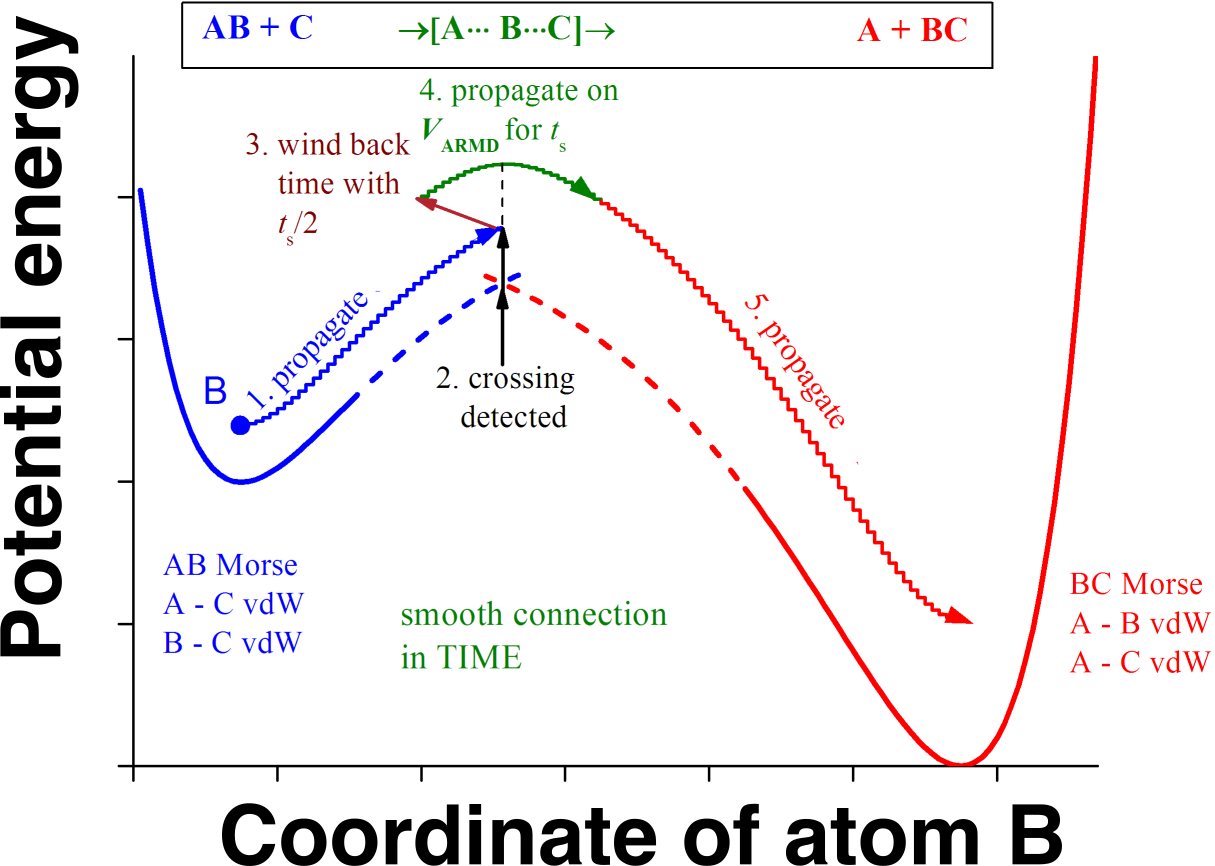}} 
\subfloat[]{\includegraphics[width = 3.1 in]{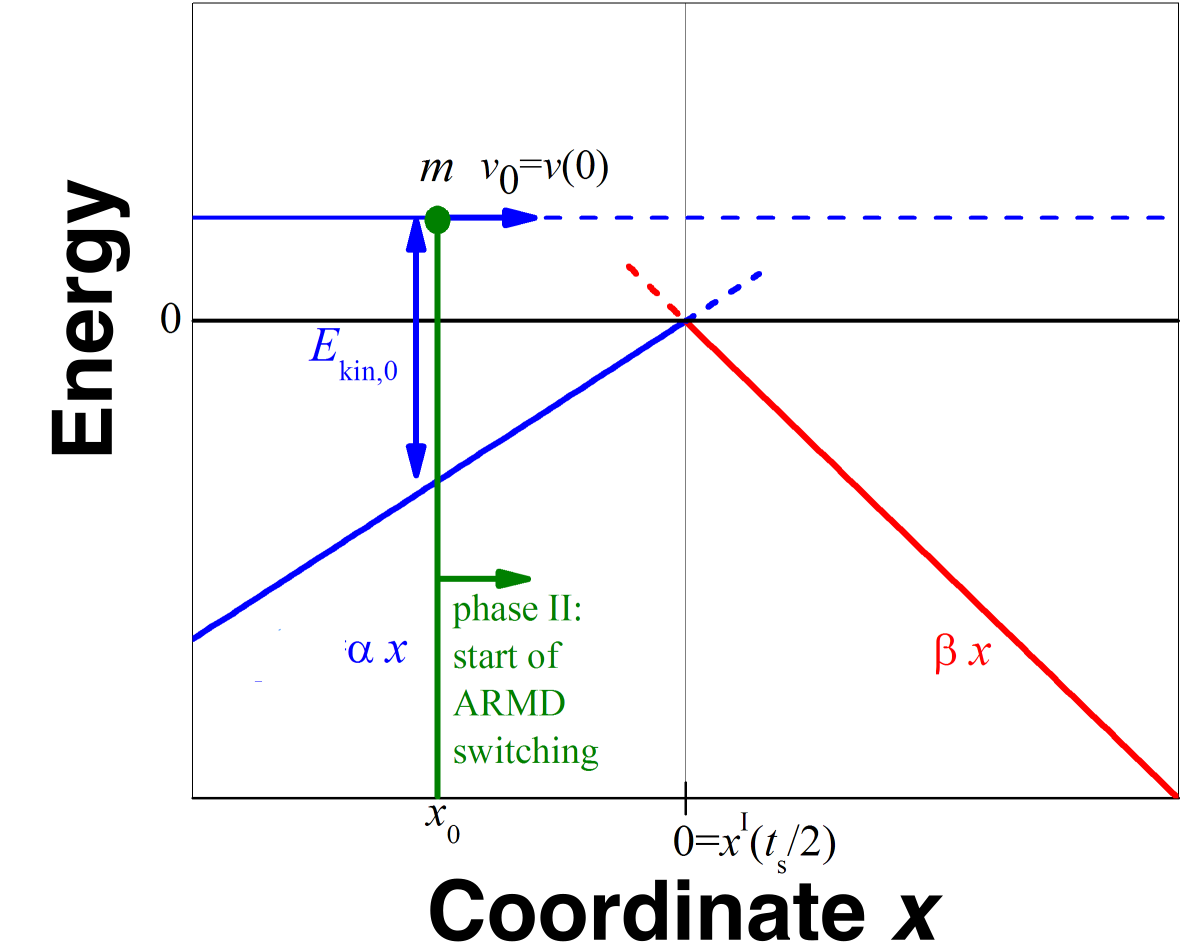}} 
\caption{a) Schematic illustration of the ARMD method for a collinear
  reaction, where atom B is transferred from donor atom A to acceptor
  atom C. During crossing the surfaces are switched in time and the
  Morse bond is replaced by van der Waals (vdW) interactions and vice
  versa. b) Simple model for estimating energy violation in ARMD
  simulations.  The system with mass $m$ approaches from the left on
  PES $V_{\rm R}(x) = \alpha x$ (phase I). At time $t=0$ it is at
  $x_0$ with velocity $v_0$ and kinetic energy $E_{\rm kin,0}$. After
  crossing is detected at $x=0$ the time is rewound by $t_{\rm s}/2$
  and the dynamics is re-simulated while $V_{\rm R}(x)$ is being
  switched to $V_{\rm P} (x) = \beta x$ in $t_{\rm s}$ (phase II). }
\label{fig:fig1}
\end{figure}

\noindent
Time resolved experiments have contributed considerably to our
understanding of chemical reactivity over the past 3 decades. Short
laser pulses (``femtochemistry'') allow to follow chemical
transformations on time scales relevant to the actual chemical step
(bond breaking or bond formation). Typical examples are the time
resolved studies of ligand (re)binding in myoglobin
(Mb)\cite{Petrich1991,Kruglik2010,lim2012mbno,negrerie2012mbno} or
vibrationally induced reactivity.\cite{Vaida:2003,crim:2009,beck:2014}
This prompted the development of time-based reactive
MD\cite{MM06cross,rmd08} whereby a chemical reaction is followed in
time which is the natural progression coordinate in such an
experiment. For the example of two states (``R'' and ``P'') the two
PESs are mixed according to $V_{\rm eff}(\vec{x}) = f(t) V_{\rm R}
(\vec{x}) + (1-f(t)) V_{\rm P} (\vec{x})$ where $f(t)$ is a sigmoid
switching function which changes from 1 to 0 between $t = -t_{\rm
  s}/2$ and $t = t_{\rm s}/2$. At the beginning of the mixing the
system is fully in the R-state ($f(t = -t_{\rm s}/2) = 1$), while at
the end it is fully in the P-state ($f(t > t_{\rm s}/2)=1$). Here, the
only free variable is the switching time $t_s$. The algorithm of ARMD
is schematically shown for a collinear atom transfer reaction in
Figure \ref{fig:fig1}a.\\

\noindent
During the mixing the system is propagated under a time-dependent
Hamiltonian which does not strictly conserve total
energy.\cite{yosa:2011,nagy.jctc.2014.msarmd} The magnitude of this
effect can be shown to scale with $1/m$ where $m$ is the reduced mass
involved in the reaction\cite{nagy.jctc.2014.msarmd} which is
inconsequential for heavy systems such as NO rebinding to Mb but
affects the dynamics in the product channel for proton transfer
processes. This led to the development of multi state adiabatic
reactive MD (MS-ARMD) which mixes the PESs with energy dependent
weights and strictly conserves energy.\cite{nagy.jctc.2014.msarmd}\\

\noindent
In MS-ARMD the PESs are mixed in energy-space according to
\begin{equation}
V_\textrm{MS-ARMD}(\vec{x}) = \sum_{i=1}^n w_i(\vec{x}) V_i(\vec{x})
\label{E_msarmd} 
\end{equation}

\begin{figure}[th]
\subfloat{\includegraphics[width = 3.05 in]{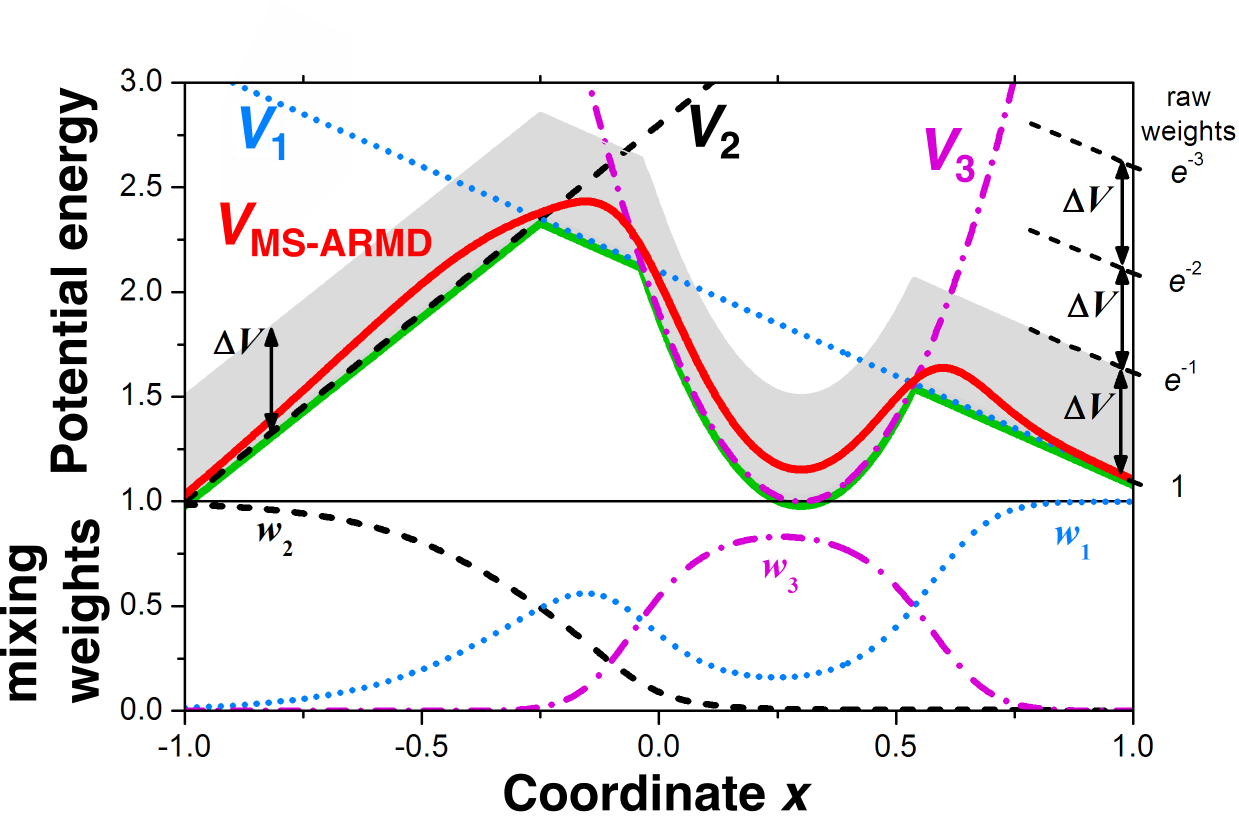}}
\subfloat{\includegraphics[width = 3.45 in]{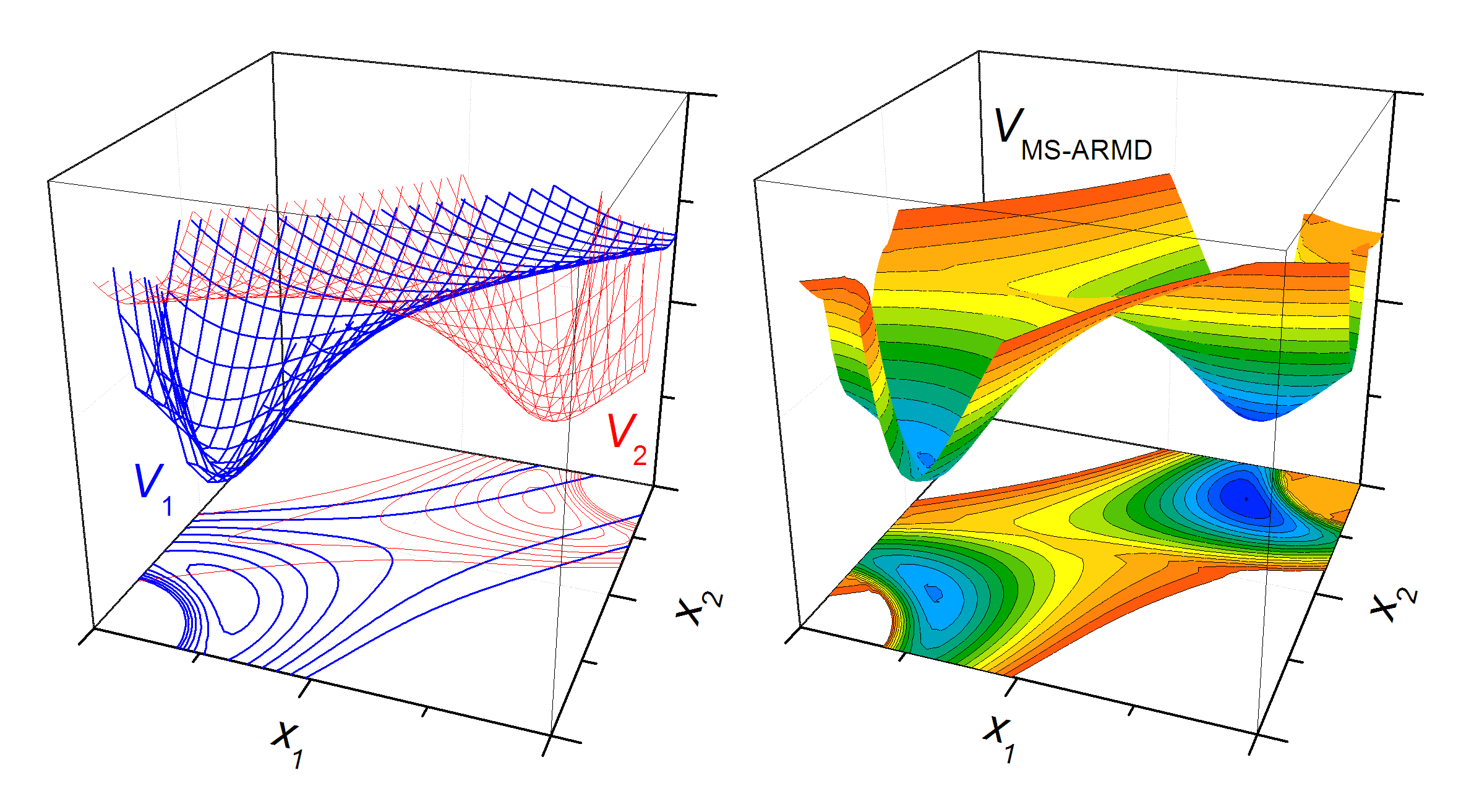}}
\caption{The MS-ARMD switching method applied in one and two
  dimensions to 3 and 2 surfaces ($V_{1,2,3}$). The effective surface
  is ($V_\textrm{MS-ARMD}$) always close to the lowest-energy surface
  ($V_\textrm{min}$), except for regions where other surfaces are
  within a few times $\Delta V$ (here $=0.5$) in energy. Here, the
  algorithm switches smoothly among them by varying their weights
  ($w_{1,2,3}$; lower left panel)}
\label{fig:fig3_msarmd1d}
\end{figure}

The weights (see Figure \ref{fig:fig3_msarmd1d}) $w_i(\vec{x})$ are
obtained by renormalizing the raw weights $w_{i,0}(\vec{x})$
\begin{equation}
w_i(\vec{x}) = \frac{w_{i,0}(\vec{x})}{\sum\limits_{ i=1}^n
  w_{i,0}(\vec{x})} \quad \quad \quad \quad {\rm where} \quad
w_{i,0}(\vec{x}) = \exp \left(-\frac{(V_i(\vec{x})-V_{\rm
    min}(\vec{x}))}{\Delta V}\right)
\label{raw_weight}
\end{equation}
where $V_{\rm min}(\vec{x})$ is the minimal energy for a given
configuration $\vec{x}$ and $\Delta V$ is a parameter. The raw weights
(Eq. \ref{raw_weight}) depend exponentially on the energy difference
between surface $i$ and the minimum energy surface $V_{\rm
  min}(\vec{x})$ over a characteristic energy scale $\Delta V$
(switching parameter). Only those surfaces will have significant
weights, whose energy is within a few times of $\Delta V$ from the
lowest energy surface $V_{\rm min}(\vec{x})$ for instantaneous
configuration $\vec{x} $. The performance of MS-ARMD is demonstrated
for crossings of 1D and 2D surfaces in Figure \ref{fig:fig3_msarmd1d}.
Finally, ARMD with energy-dependent weights mixes the different PESs
by using Gaussian and polynomial functions (GAPOs) in the neighborhood
of crossing points between the states. The parameters of these GAPOs
need to be determined through fitting to reference data (e.g. from a
calculation along the intrinsic reaction coordinate (IRC)). This step
is the most demanding part in MS-ARMD.\cite{nagy.jctc.2014.msarmd} A
smooth global surface is obtained everywhere, even for energies where
more than two surfaces approach one another. Because the mixed PES
$V_{\rm MS-ARMD}(\vec{x})$ depends on the energies of the different
states through the weights $w_i$ which in turn are analytical
functions of the coordinates $\vec{x}$, the derivatives can be readily
determined which leads to energy conservation in MD simulations.\\

\noindent
A recent extension\cite{mm.msvalb:2018} of MS-ARMD is its combination
with VALBOND, a force field that allows to describe the geometries and
dynamics of metal complexes.\cite{Landis1998a,Firman2001,valbond09}
Here, the formulation is reminiscent of EVB whereby the diagonal terms
are VALBOND descriptions of the states involved and the off-diagonal
elements describe the orbital overlap. Furthermore, MS-ARMD can also
be combined with molecular mechanics with proton transfer
(MMPT)\cite{mmpt08} to follow proton transfer in the gas- and
condensed
phase.\cite{MM.fad:2016,MM.oxa:2017,review.kie:2017,mm.msmmpt:2019}\\

\section{Applications in Vibrational Spectroscopy}
Vibrational spectroscopy is a sensitive tool for quantitatively
characterizing the structure of and the structural dynamics around a
solute embedded in an environment.\cite{hamm:1999} Such studies can
also provide important insight into the dynamical coupling between the
solute and the solvent. One example is the possibility to infer ligand
binding strengths from infrared spectroscopic
measurements.\cite{boxer:2006} A quantitative study for benzonitrile
in lysozyme has been carried out recently using atomistic simulations
with non-conventional force fields.\cite{MM.bzn:2017}\\

\noindent
Refined multipolar force fields also allow quantitative comparisons
with experiments and their interpretation. One of the noticeable
examples is the infrared spectrum of photodissociated carbon monoxide
(CO) in myoglobin (Mb).\cite{Lim95b} The strong (43 MV/cm
\cite{Boxer99}) inhomogeneous electric field in the heme pocket leads
to characteristic shifting and splitting of the spectral lines due to
the Stark effect. Several attempts were made
\cite{Straub97,Elber98,Anselmi07} to correctly interpret the
experimental infrared spectrum using computational methods. Although
some of them were capable of correctly modelling the width of the
spectrum they usually were unable to find the characteristic splitting
of the CO spectrum (i.e., $\approx 10$ cm$^{-1}$). A first successful
attempt used a fluctuating point-charge model\cite{nutt-biophys-03}
based on an earlier three-point model for CO \cite{Str91}. This was
later generalized to a rigorous fluctuating MTP model which reproduced
most features of the spectrum known from experiments.\cite{Plattner08}
In particular, the splitting, width and relative intensities of the
computed spectrum agreed favourably with the experimentally known
properties. Based on this it was then also possible to assign the two
spectroscopic signatures to distinct conformational substates. Those
agreed with results from more indirect (based on mutations in the
active site)\cite{Nienhaus:2005} or more difficult-to-converge mixed
QM/MM simulations from MD simulations.\cite{MM.cpc:2006}\\

\noindent
Understanding structural, spectroscopic and dynamical properties of
water in its different phases provides a rich variety for developing
and improving special-purpose, tailored and physics-based force
fields. Among the most successful for spectroscopic applications are
MB-pol\cite{mbpol:2013} and WHBB.\cite{bowman.whbb:2011} Other,
similarly successful models, include the explicit 3-body (E3B)
parametrization,\cite{e3b:2008} the (i)AMOEBA family of force
fields,\cite{amoeba:2003,iamoeba:2013} the TL$n$P
family\cite{tavan:2013} and a recent modification of
it.\cite{MM.water:2018}\\

\noindent
The ManyBody-polarizable (MB-pol) model is based on two- and
three-body water interactions calculated at the CCSD(T) level of
theory. It is entirely developed from
first-principles.\cite{mbpol:2013,mbpol:2014} MB-pol explicitly treats
the one-body (monomer distortion energy) term and the short-ranged
two- and three-body terms akin to a polarizable potential supplemented
by short-range two- and three-body terms that effectively represent
quantum-mechanical interactions arising from the overlap of the
monomer electron densities. Specifically, at all separations, the
total MB-pol two-body term includes (damped) dispersion forces derived
from ab initio computed asymptotic expansions of the dispersion energy
along with electrostatic contributions due to the interactions between
the molecular permanent and induced moments. Similarly, the MB-pol
three-body term includes a three-body polarization term at all
separations, which is supplemented by a short-range 4th-degree
permutationally invariant polynomial that effectively corrects for the
deficiencies of a purely classical representation of the three-body
interactions in regions where the electron densities of the three
monomers overlap. This short-range three-body contribution is smoothly
switched off once the oxygen-oxygen separation between any water
molecule and the other two water molecules of a trimer reaches a value
of 4.5 Å. In MB-pol, all induced interactions are described through
many-body polarization. MB-pol thus contains many-body effects at all
monomer separations as well as at all orders, in an explicit way up to
the third order and in a mean-field fashion at all higher
orders. Using MD simulations the computed infrared and Raman spectra
for liquid water can be realistically modelled. Finally, molecular
dynamics (MB-MD) simulations carried out with MB-pol describe the
infrared (IR) and Raman spectra of liquid water well compared with
experiments.\cite{paesani:2015}\\

\noindent
The WHBB model\cite{bowman.whbb:2011} is the sum of monomer, 2-body,
and 3-body, full-dimensional potentials, with the option of including
4 and higher-body interactions from other polarizable potentials. The
monomer PES is spectroscopically accurate and the 2- and 3-body
potentials are mathematical fits to tens of thousand of ab initio
electronic energies (CCSD(T)/aug-cc-pVTZ (aVTZ) and MP2/ aVTZ
energies) using permutationally invariant
polynomials.\cite{bowman.irpc:2009} The WHBB model has been applied to
the infrared spectroscopy of liquid water and showed to perform well
compared with experiment.\cite{bowman.whbb:2015}\\

\begin{figure}[th]
\subfloat[]{\includegraphics[width = 3.20 in]{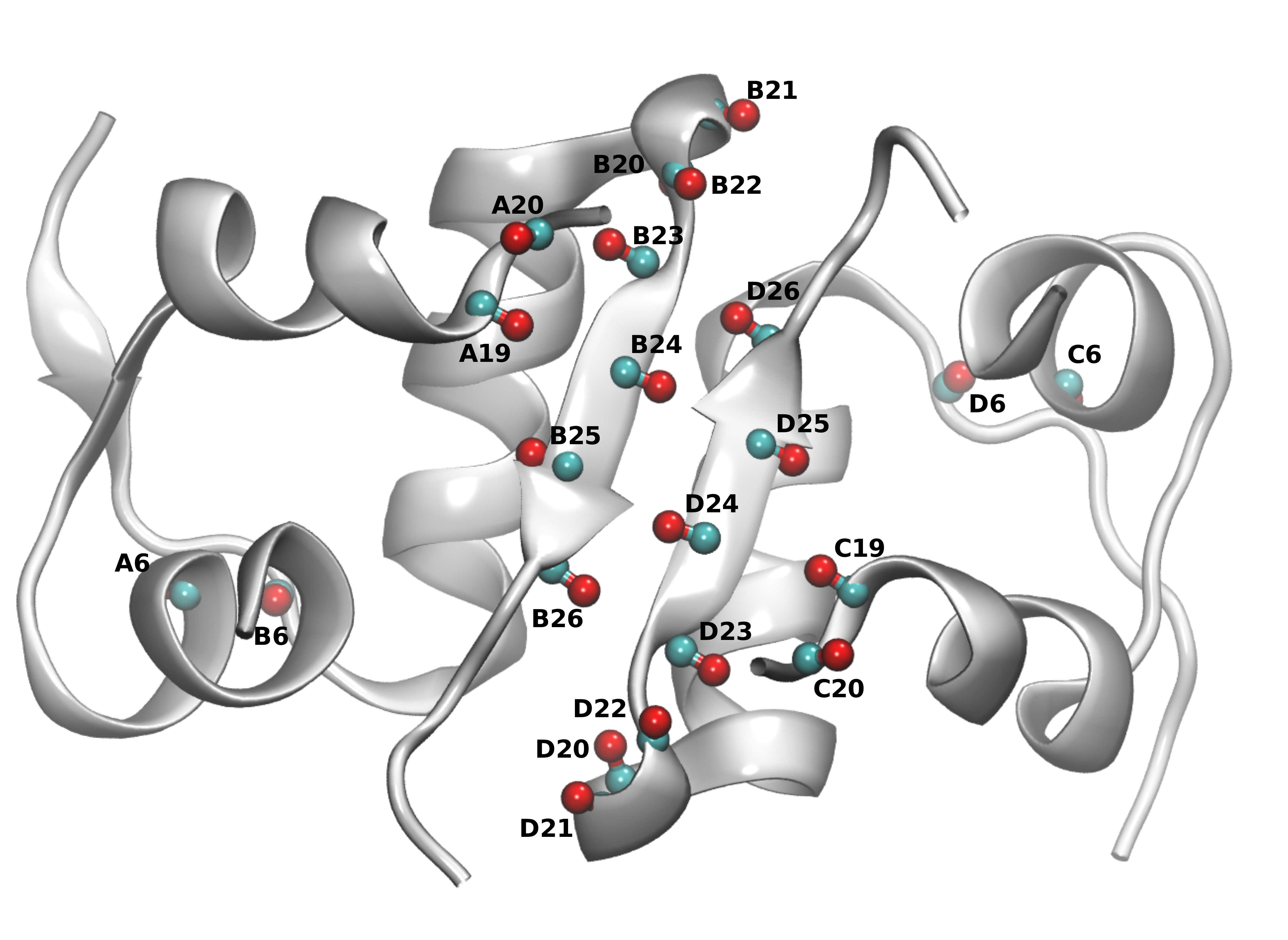}} 
\subfloat[]{\includegraphics[width = 3.20in]{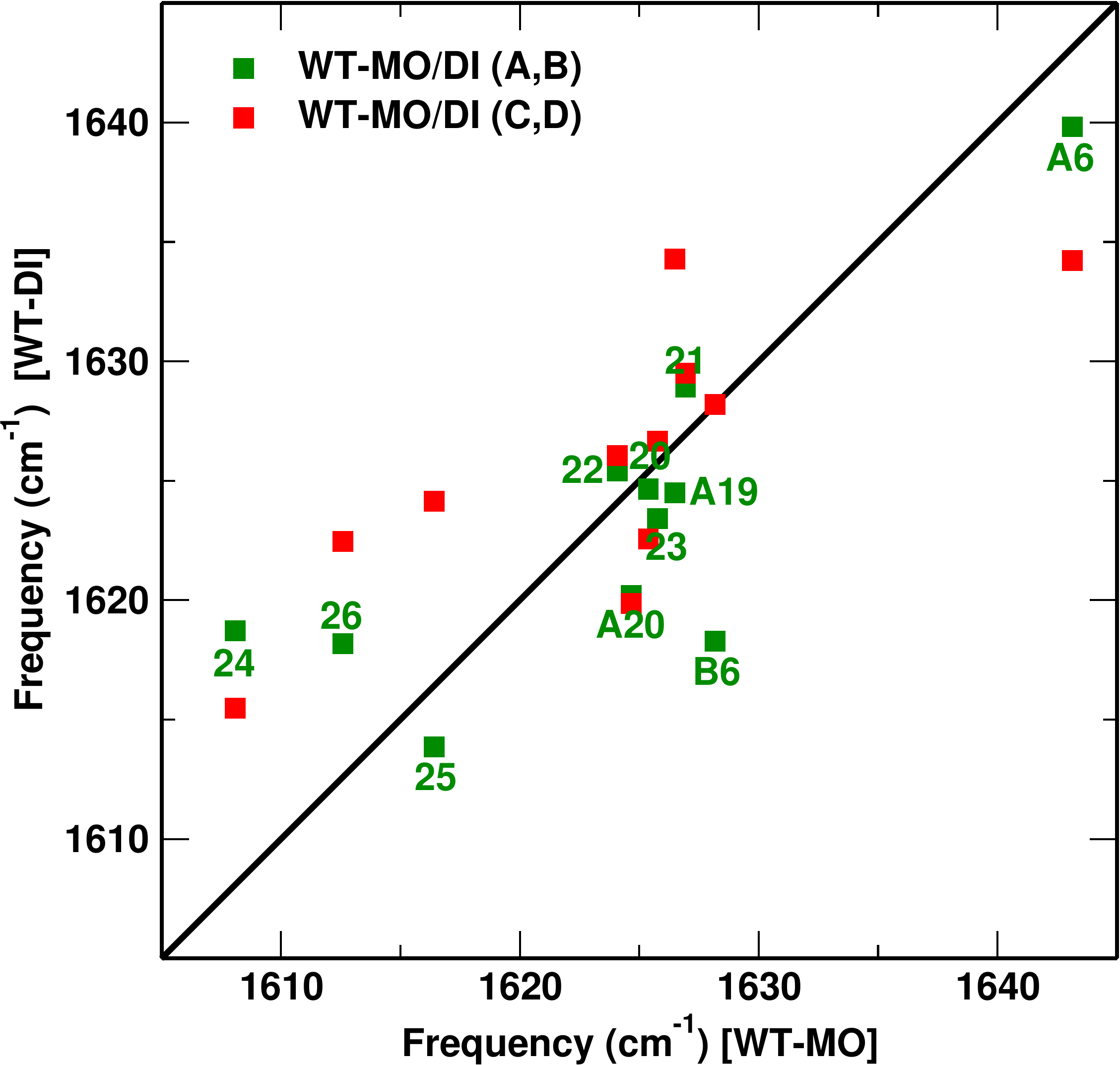}} 
\caption{a) Structure of the WT insulin dimer with the CO labels
  considered specifically. b) Comparison between the maxima of the
  frequency distribution for WT-MO and WT-DI. Residues of chain A/B
  and C/D are shown in green and red, respectively.}
\label{fig:insulin}
\end{figure}

\noindent
Multipolar force fields have recently been used for a number of 1d-
and 2d-spectroscopy studies including
N-methylacetamide\cite{MM.nma:2014},
fluoro-acetonitrile\cite{MM.facn:2015}, cyanide\cite{MM.cn:2013},
insulin in water\cite{MM.insulin:2019}, and azide in gas phase and in
water.\cite{MM.n3:2019} As an example for using such simulations in a
concrete context, the 1d-infrared spectroscopy of wild type insulin
monomer (WT-MO, chains A and B, see Figure \ref{fig:insulin}) and
dimer (WT-DI, chains A to D) in solution is discussed here. For this,
the X-ray crystal structure of WT insulin dimer resolved at 1.5 \AA\/
(protein data bank (PDB\cite{Bernstein.pdb:1977,Berman.pdb:2002})
code: 4INS)\cite{Baker.pdb:1988} were solvated in a cubic box ($75^3$
\AA\/$^3$) of TIP3P\cite{Jorgensen.tip3p:1983} water molecules. For
all molecular dynamics (MD) simulations the
CHARMM\cite{Brooks.charmm:2009} package together with the
CHARMM36\cite{charmm36:2010} force fields was used. The systems were
minimized and equilibrated for 20 ps in the $NVT$ ensemble. Production
runs, 1 ns in length, were carried out in the $NPT$ ensemble, with
coordinates saved every 10 fs for subsequent analysis. A velocity
Verlet\cite{Swope.vv:1982} integrator and Nos\'{e}-Hoover
thermostat\cite{Nose:1984,Hoover:1985} was employed in the $NVT$
simulations. For the $NPT$ simulations, an Andersen and
Nos\'{e}-Hoover constant pressure and temperature
algorithm\cite{Andersen:1980,Nose:1983,Hoover:1985} was used together
with a leapfrog integrator.\cite{Hairer.lpf:2003}\\

\noindent
Time-ordered snapshots were extracted from the MD simulation every 10
fs and then the anharmonic frequencies were computed. For this
purpose, one-dimensional potential energy curves were calculated along
the -CO normal modes for each snapshot. The anharmonic transition
frequencies ($v = 0 \rightarrow$ 1) were then calculated by solving
the 1D time-independent Schr\"odinger equation to obtain the frequency
trajectory. Figure \ref{fig:insulin}B compares the maximum of the
frequency distribution for $10^5$ snapshots between monomer and dimer
for several residues (6, 19, and 20 from chains A/C and 6, and 19-26
from chains B/D) for the WT protein. Residues 24 to 26 of the B/D
chains are involved in hydrogen bonding at the dimerization
interface. As is shown in Figure \ref{fig:insulin}B, the maximum
frequencies for residues 24 and 26 of chain B/D are both blue shifted
when going from the monomer to the dimer while for residue 25B a small
red shift is observed and clearly blue shifted for residue 25D. This
is also consistent with previous results\cite{MM.insulin:2019} which
showed that the IR spectra of the symmetry-related pairs of chain B/D
do not superimpose. This also suggests that the structural dynamics
and spectroscopy of the CO probes involved in the protein-protein
contact (24-26 B/D) represent characteristic changes and this feature
can be used to distinguish the monomeric from dimeric insulin in
solution.  \\

\noindent
As a second example, the influence of the scanning direction in
determining the local frequency is explored. For this, simulations of
N-methyl acetamide (NMA) were carried out in which the entire NMA
molecule was treated with MTP
electrostatics.\cite{MM.mtp:2012,MM.mtp:2013,MM.nma:2014,mm.nma:2015}
For the bonded terms, a reproducing kernel Hilbert space (RKHS-)based
representation of the PES has been constructed for the amide-I stretch
(CONH) fragment. For convenience, this was based on PBEPBE/cc-pVTZ
reference calculations although the approach can also be applied to
reference data calculated at much higher levels of theory. In the
simulations, the electrostatic interactions are treated using
Particle-Mesh Ewald\cite{Darden.pme:1993} (PME) with grid size spacing
of 1 \AA\/, characteristic reciprocal length $\kappa = 0.32$
\AA\/$^{-1}$, and interpolation order 4. All bonds involving hydrogen
atoms were constrained using SHAKE\cite{Gunsteren.shake:1997} except
for the NH bond in amide-I mode .\\

\noindent
To construct the RKHS PES (for details on RKHS,
see\cite{rabitz:1996,MM.rkhs:2017}), the NMA structure was optimized
first and then the normal modes are calculated by fixing all hydrogen
atoms of the methyl groups. Next, gas phase PBEPBE/cc-pVTZ energies
and forces were calculated using Gaussian09 software\cite{g09} for
2400 structures sampled randomly along the 12 normal modes.  In order
to describe the PES for the NMA fragment (X(CONH)Y where ``X'' and
``Y'' are the two methyl groups as united atoms), a multidimensional
reproducing kernel based function is considered which is the sum of 15
four-body interactions ($^6C_4 = 15$ where $^nC_k$ is the binomial
coefficient). Each of the four-body interactions is represented by a
6-dimensional tensor product kernel for 6 radial dimensions (the
interatomic distances in a tetra-atomic systems). A reciprocal power
decay kernel of type $(n=3,m=0)$ was used for all the bond
distances.\cite{rabitz:1996,MM.rkhs:2017} The kernel coefficients were
calculated using least square fitting. Finally RKHS PES has been
constructed for the Amide-I stretch mode. This RKHS PES describes all
relevant bonded terms for the CONH motif and also the bonded
interaction in H$_3$C-C and H$_3C$-N. The harmonic frequency
calculated for the Amide-I stretch from the RKHS PES compares well
with the one obtained from the DFT calculations (1696 vs. 1698
cm$^{-1}$) which serves as a validation of the present approach.\\

\noindent
To simulate the dynamics of NMA, the RKHS-based representation for the
bonded terms is implemented in CHARMM and supplemented with
conventional force field parameters to describe the interactions
between the methyl hydrogen atoms and all other atoms. The harmonic
frequency of the amide-I stretch for NMA is at 1714 cm$^{-1}$,
slightly higher than that computed using the stand-alone RKHS PES
(1696 cm$^{-1}$) from above. This is due to coupling between the CONH
motif and the rest of the NMA molecule. Next, MD simulations are run
for the solvated system and the 1-dimensional potential energy is
scanned along two different directions: the CO and the CONH
directions, see Figure \ref{fig:nma-water-ffcf}, to investigate the
differences depending on the approach chosen.\\

\begin{figure}
\includegraphics[width = 4.05 in]{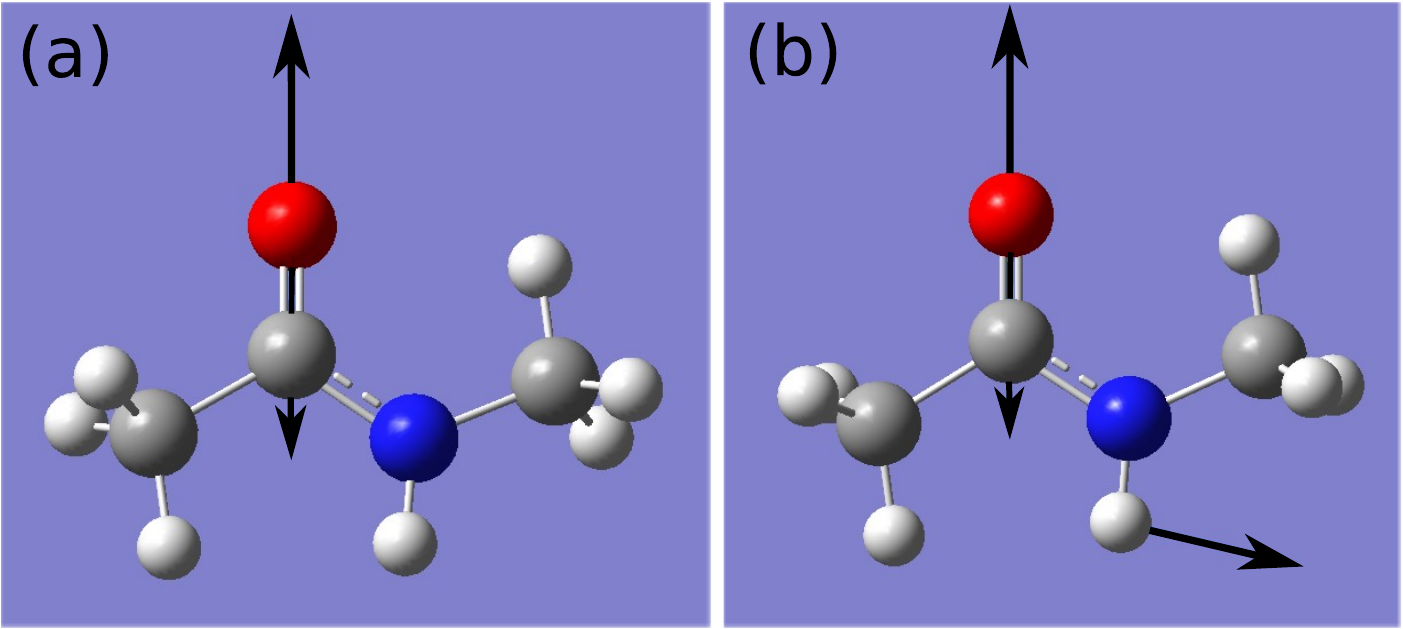}
\caption{Displacement vector for two different scanning to construct
  an 1D potential. (a) Distortion along the CO stretching and (b)
  movement along the CONH stretching.}
\label{fig:nma-modes}
\end{figure}

\noindent
The anharmonic frequencies for the fundamental and overtones of a
particular mode and corresponding eigen functions can be calculated by
solving 1D Schr\"{o}dinger equation using a discrete variable
representation (DVR) approach.\cite{col92:1982} For the amide-I mode
in NMA - which can also be applied to the protein backbone, see above
- two different schemes are followed: (a) scanning along the CO
stretch, and (b) displacing the CONH group along the amide-I normal
mode (see Figure \ref{fig:nma-modes}). The potential energies are
again represented by a 1d-reproducing kernel and the Schr\"{o}dinger
equation is then solved using a DVR approach for $-\infty < r <
\infty$, where $r$ is the displacement from the equilibrium
geometry. The anharmonic stretching frequencies in the gas phase are
computed as 1671 and 1695 cm$^{-1}$ using the potential obtained
following schemes (a) and (b), respectively. The reduced mass is 1 amu
for both cases, since unnormalized displacement coordinates are used
to distort the molecule along the normal modes.\\

\begin{figure}
\includegraphics[width = 6.05 in]{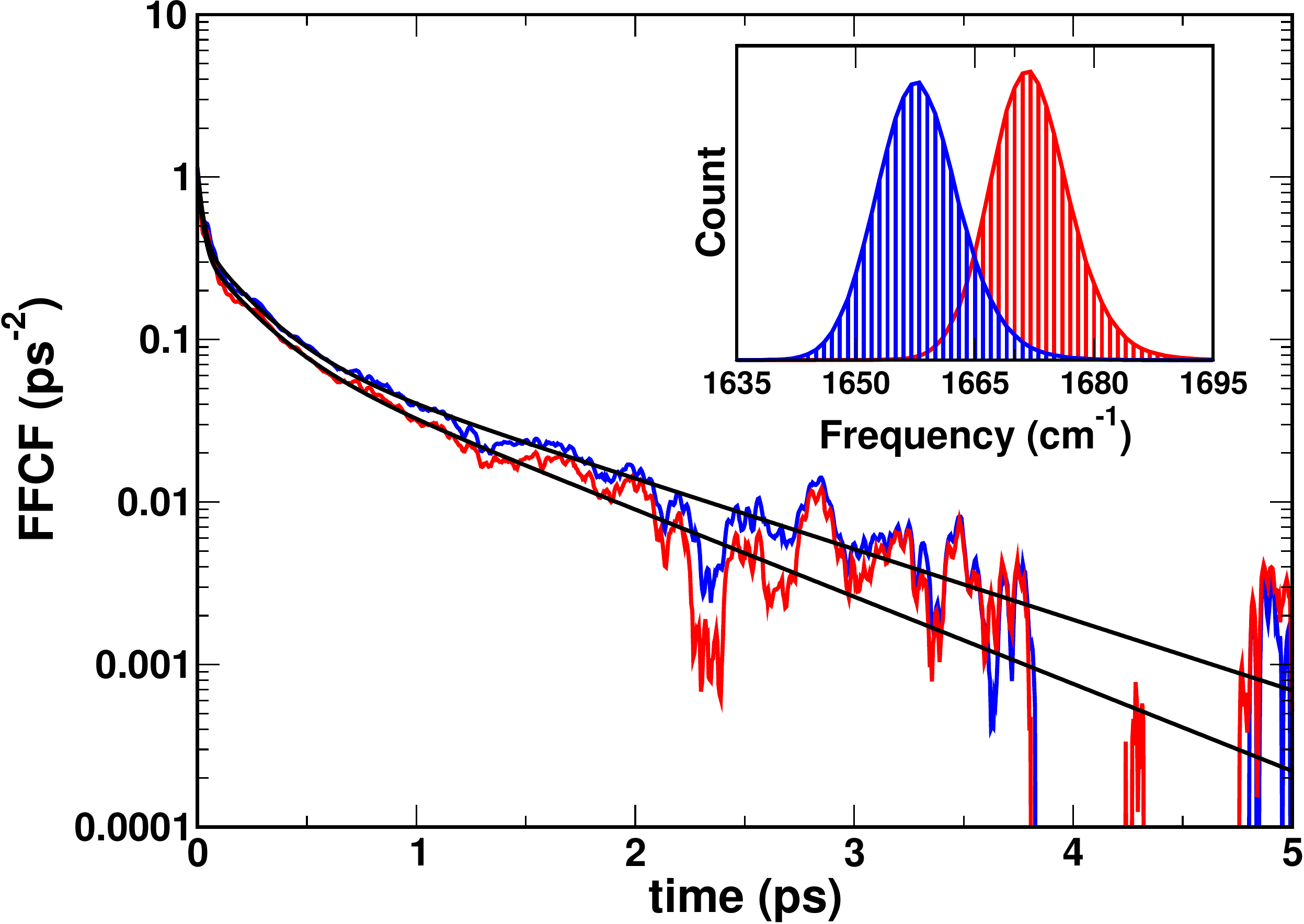}
\caption{Frequency fluctuation correlation functions (FFCF) of NMA in
  water with anharmonic frequency calculated via scanning along CONH
  mode (red) and CO stretching mode (blue) using multipolar force
  fields and gas phase RKHS potential of NMA described above. The
  black lines refer to the tri-exponential fit to the FFCF. Inset
  shows the corresponding frequency distributions.}
\label{fig:nma-water-ffcf}
\end{figure}

\noindent
Using the RKHS based PES described above, a 4 ns trajectory of
solvated NMA (30$^3$ \AA\/$^3$ TIP3P water\cite{Jorgensen.tip3p:1983})
in the $NVE$ ensemble was run following 20 ps of heating and 1.2 ns of
equilibration at 300 K. The trajectory was saved every 5 fs and the
anharmonic frequencies were determined from scanning the
RKHS-potential along the CO and the CONH directions (see Figure
\ref{fig:nma-modes}) for $8 \times 10^5$ snapshots using MTP
electrostatics. The MTPs mainly account for the NMA-water
electrostatic interactions while the gas phase RKHS potential
accurately represents the amide-I mode within NMA. Figure
\ref{fig:nma-water-ffcf} shows the decay of the frequency fluctuation
correlation function (FFCF) and frequency distribution (see inset in
Figure \ref{fig:nma-water-ffcf}) for NMA in water with anharmonic
frequency calculated via scanning along the two directions.  The
average frequencies for scanning along the CO (a) and CONH mode (b) of
NMA in water are at 1657 cm$^{-1}$ and 1671 cm$^{-1}$, respectively. A
tri-exponential fit to the frequency fluctuation correlation functions
yields three timescales which are close to one another from approaches
(a) and (b) and compare quite well with
experiment\cite{Woutersen:2002,Decamp:2005} and previous
simulations,\cite{MM.nma:2014} see Table \ref{tab:tab1}.\\

\begin{table}[H]
\centering
\begin{tabular*}{\linewidth}{@{\extracolsep{\fill}} lcccccc}
\hline
Method & $a_1$ [ps$^{-2}$] & $t_1$ [ps] & $a_2$ [ps$^{-2}$] & $t_2$ [ps] & $a_3$ [ps$^{-2}$] & $t_3$ [ps] \\
\hline
CO scan  & 0.71 & 0.021 & 0.302 & 0.208 & 0.103 & 1.0  \\
CONH scan & 0.70 & 0.020 & 0.25 & 0.20 & 0.106 & 0.81 \\
\hline
sim. \cite{Decamp:2005} &  &  0.06 & && & 0.66 \\
exp. \cite{Woutersen:2002} &  &  (0.05-0.1) && & & 1.6 \\
exp. \cite{Decamp:2005} &  &  0.01 & & && 1.0 \\
\hline
\end{tabular*}
\caption{The FFCF decay times and their amplitudes for scanning along
  the CO and CONH normal modes for solvated NMA obtained from a
  tri-exponential fit, see Figure \ref{fig:nma-water-ffcf}. The
  frequency calculations used the RKHS PES and MTP electrostatics. For
  comparison, vibrational decay times from
  experiment\cite{Woutersen:2002} and simulations\cite{Decamp:2005}
  are also given.}
\label{tab:tab1}
\end{table}

\noindent
As a final example for a spectroscopic application, a recent model for
azide in gas phase and in solution is discussed. The bonded terms were
again based on RKHS which exactly reproduces the energies from the
reference {\it ab initio} calculations. For N$_3^-$, a 3-dimensional
PES based in multi reference CI (MRCI) calculations was represented as
an RKHS and quantum bound states were determined.\cite{MM.n3:2019} The
agreement with gas phase experiments was found to be good, typically
within $\sim 10$ cm$^{-1}$. Simulations in solution provided the
correct blue shift, consistent with experiment. This opens now the
possibility to use the new parametrization of N$_3^-$ for simulations
of proteins in solution and to determine the site-specific dynamics
depending on the position of the label.\\

\section{Applications in Reaction Dynamics}
The investigation of chemical reactions using atomistic simulations
dates back to the 60s when the first studies of the H+H$_2$ reactions
were carried out.\cite{karplus:1964} One unexpected finding was that
such quasi-classical trajectory (QCT) simulations agree quite well
with quantum simulations\cite{schatz:1976} despite the fact that the
H+H$_2$ system is particularly susceptible to quantum effects
including zero point energy and tunneling. In the following, QCT
simulations became an established way to investigate small- and
large-molecule reactive processes.\\

\subsection{Small-Molecule Reactions Involving Tri- and Tetra-Atomics}
The field of small molecule reactions was heavily driven by the advent
of refined experimental techniques, such as molecular beam studies or
photodissociation spectroscopy which provide energy-, state- and
conformationally-resolved information about reactive molecular
collisions. Initially, empirical energy functions, such as the
London-Eyring-Polanyi-Sato (LEPS) surface,\cite{london29,lep31,leps2}
were used to carry out QCT simulations. Once electronic structure
calculations became sufficiently accurate and the configurational
space could be covered by computing energies for many geometries, such
empirical PESs were largely superseded. The primary problem then
shifted to representing the computed points such that the total
potential energy can be evaluated with comparable accuracy as the
underlying quantum chemical calculations.\\

\noindent
For triatomic molecules a convenient representation was one that used
an expansion into Legendre polynomials $P_{\lambda}(\cos{\theta})$ and
corresponding radial strength functions $V_{\lambda}(R,r)$,
i.e. $V(R,r,\theta) = \sum_{\lambda=0}^{\lambda_{\rm max}}
V_{\lambda}(R,r) P_{\lambda}(\cos{\theta})$. Such expansions have been
particularly useful for studying van der Waals
complexes.\cite{H90Ann,H91Adv,MM99:heh2p,MM99:nehf} Such
representations have been used either together with explicit fits to
experimental data, or by fitting them to reference electronic
structure calculations. An alternative is to separate the total
interaction energy into long- and short-range parts and to represent
them separately. Often, the long range part in such complexes can be
described very accurately by resorting to classical electrostatics
based on atomic and molecular multipoles and polarizabilities and
hyperpolarizabilities.\cite{avoird:2017,MM.heh2:2019}\\

\noindent
An alternative is an expansion into permutationally invariant
polynomials (PIPs)\cite{bowman.irpc:2009} which has, for example, been
applied to N$_4$ for studying reactive collisions for N$_2$ + N$_2$
$\rightarrow$ N$_2$ + 2N and N$_2$ + N$_2$ $\rightarrow$
4N.\cite{candler:2014} PIPs employ a basis of exponential or
Morse-type functions to expand the PES and fit products of such basis
functions to the reference data. The permutational symmetry of
chemically identical atoms is preserved in such an
approach. Similarly, a study of N$_4^+$ using an RKHS representation
also incorporated permutational invariance
explicitly.\cite{MM.n4:2012}\\

\noindent
Another method is a representation based on a Reproducing Kernel
Hilbert Space
(RKHS).\cite{rabitz:1996,hollebeek.annrevphychem.1999.rkhs,MM.rkhs:2017}
Such an approach {\it exactly} reproduces the input data which are
usually results from electronic structure calculations on a predefined
grid. One of the advantageous features of an RKHS is that the physical
long range decay for a radial coordinate $\rightarrow \infty$ can be
encoded in the functional form of the kernel. Such RKHS-based
representations have been used for entire PESs\cite{MM.cno:2018} or in
a QM/MM-type approach to treat part of an extended system with higher
accuracy.\cite{MM.cco:2013,meuwly.2016.mbno,MM.trhbn:2018}\\

\noindent
Recently, RKHS-based reactive PESs have been successfully used for a
range of atom+diatom reactions. For these systems the focus was
primarily on thermal reaction rates, final state distributions and
vibrational relaxation times which are properties of particular
interest to hypersonics and atmospheric
re-entry.\cite{sarma:2000,cummings:2003,dsmc:2017} Reference data for
the systems of interest ([NOO], i.e. N+O$_2$ and O+NO; [NNO], and
[CNO]) can be determined at high levels of theory (MRCI) with
good-quality basis sets (aug-cc-pVTZ or
larger).\cite{MM.no2:2020,alp17:2392,MM.n2o:2020,MM.cno:2018} A
typical PES for one electronic state is based on $~10^4$ energies and
the RKHS-interpolated surfaces reproduce the reference data to within
much better than chemical accuracy, typically within a few
cm$^{-1}$. Extensive QCT simulations on these PESs now provide a
comprehensive and consistent set of thermal rates and vibrational
relaxation times for the most relevant systems involved in
hypersonics.\cite{MM.hyperson:2020}\\

\subsection{Larger Molecules and Proteins}
\noindent
For larger molecules in the gas phase, MS-ARMD has also been used
together with established empirical force fields. This has been
applied to the study of specific reaction channels such as proton
transfer within H$_2$SO$_4$ or photodissociation of H$_2$SO$_4$
$\rightarrow$ H$_2$O + SO$_3$\cite{reyes.pccp.2014.msarmd} and other
atmospherically relevant molecules following vibrational excitation of
the OH stretch\cite{reyesbrickel.pccp.2016.msarmd,MM.hso3f:2017}, the
Claisen rearrangement reaction\cite{MM.claisen:2019} or to investigate
Diels-Alder reactions.\cite{MM.diels:2019} Such studies provide
molecular-level details into the reaction mechanisms and relevant
coordinates driving the reaction based on dynamics studies which goes
beyond scanning and relaxing the energy function along a minimum
energy path. As an example, for the Diels-Alder reaction between
2,3-dibromo-1,3-butadiene and maleic anhydride such reactive dynamics
simulations emphasized the important role played by rotations or the
two reactants to reach the transition state.\cite{MM.diels:2019} \\

\noindent
Finally, biological systems can also be studied with a combination of
RKHS-based PES and an empirical force
field.\cite{meuwly.2016.mbno,MM.trhbn:2018} Such studies allowed a
structural interpretation of metastable states in MbNO and a
molecularly refined understanding of ligand exchange (NO vs. O$_2$) at
the heme-iron in truncated hemoglobin.\\

\section{Outlook}
Very recent developments in accurate representations of
multidimensional potential energy surfaces use generalizations of
PIPs\cite{bowman.nma:2019} or machine learning within the context of
kernel ridge regression\cite{fchl:2018} or neural
networks.\cite{ani:2017,schnet:2018,MM.physnet:2019} Common to all
these approaches is the fact that they are based on extensive
reference data sets. Also, the evaluation time of such representations
is computationally considerably more expensive than that of a
parametrized empirical function. For the Diels-Alder reaction between
2,3-dibromo-1,3-butadiene and maleic anhydride evaluating the MS-ARMD
and NN-based PESs differs by a factor of $\sim
200$.\cite{MM.diels:2019}\\

\noindent
As an example for future applications and improvement, the possibility
to run MD simulations of molecules with machine learned energy
functions and charges in solution is
mentioned. PhysNet\cite{MM.physnet:2019} can be used to train
energies, forces and partial charges which yields accurate
representations of high-dimensional, reactive
PESs.\cite{MM.diels:2019,mm.ht:2020,mm.atmos:2020} As has been noted
above, accurate representations of the ESP are required for
quantitative investigations of the vibrational spectroscopy in the
condensed phase. As PhysNet provides geometry-dependent charges in a
natural way, simulations of solutes in aqueous and other environments
will benefit greatly from such more realistic representations. This
also brings atomistic simulations based on empirical energy functions
closer to more rigorous mixed quantum/classical simulations.\\

\noindent
It is expected that with the increased performance of computer
architecture and the recent advances in machine learning and
representing high-dimensional potential energy surfaces, the quality
of simulations involving small molecules will further increase. This
will bring experiment, simulation and theory closer together and
provide new opportunities to characterize, analyze, and predict
molecular properties and processes at an atomic scale. A next step in
this endeavour is to transfer and apply this technology to larger
systems, including proteins and materials.\\

\section*{Acknowledgments}
This work was supported by the Swiss National Science Foundation
grants 200021-117810, 200020-188724, the NCCR MUST, the AFOSR, and the
University of Basel which is gratefully acknowledged.

\bibliography{literature}

\end{document}